\documentstyle[12pt,aasms4] {article}

\begin{document}

\title{Spectroscopy of Infrared Flares from the Microquasar GRS
1915+105}

\author{S.S. Eikenberry\altaffilmark{1,2}, K. Matthews, T.W. Murphy Jr.,
R.W. Nelson}

\affil{Department of Physics \\ California Institute of Technology
\\ Pasadena, CA  91125}

\author{E.H. Morgan, R.A. Remillard, M. Muno}

\affil{Center for Space Research \\ Massachusetts Institute of
Technology \\ 77 Massachusetts Avenue \\ Cambridge, MA  02139}

\altaffiltext{1}{Sherman Fairchild Postdoctoral Fellow in Physics}
\altaffiltext{2}{Current Address: Department of Astronomy, Cornell
University, Ithaca, NY  14853}

\begin{abstract}

	We present near-infrared medium-resolution ($R \sim 875$)
spectra of the microquasar GRS 1915+105 on 1997 August 13-15 UTC from
the Hale 200-inch telescope.  The spectra show broad emission lines of
He\,I (2.058 $\mu$m) and H\,I (2.166 $\mu$m - Br$\gamma$), consistent
with previous work.  On August 14 UTC, we took spectra with $\sim
6$-minute time resolution during infrared flaring events similar to
those reported in Eikenberry et al. (1998a), which appear to reveal
plasma ejection from the system.  During the flares, the emission line
fluxes varied in approximately linear proportionality to the IR
continuum flux, implying that the lines are radiatively pumped by the
flares.  We also detected a weak He\,II (2.189 $\mu$m) emission line
on August 14 UTC.  The nature of the line variability and the presence
of the He\,II feature indicate that the emission lines in GRS 1915+105
arise in an accretion disk around the compact object, rather than in
the circumstellar disk of a proposed Oe/Be companion.  The radiative
line pumping also implies that the flare emission originates from
ejecta which have moved out of the accretion disk plane.

\end{abstract}


\section{Introduction}

	The microquasar GRS 1915+105 has attracted considerable
interest since its discovery as a hard X-ray transient by GRANAT/SIGMA
in 1992 (Castro-Tirado et al., 1995), and study has intensified since
the discovery of apparently superluminal jets emanating from the
system in 1994 (Mirabel and Rodr\'{\i}guez, 1994), which made it the
first member of the class of Galactic microquasars.  The microquasars
offer tremendous potential as closer, smaller, and faster analogs of
the quasars that emit bright radio jets.  Recently, multiwavelength
studies have shown that GRS 1915+105 exhibits episodes of
quasi-periodic ($\sim 20-30$ minutes) X-ray dips, where the X-ray flux
drops by nearly an order of magnitude in a few seconds and then
abruptly recovers.  Brightness oscillations with similar timescales
have also been seen in the radio and infrared(Pooley \& Fender, 1997;
Fender et al., 1997).  Fitting the X-ray spectra during this behavior,
Belloni et al. (1997) have shown that the dips are consistent with a
scenario where the X-ray-emitting inner portion of an accretion disk
{\it disappears}.  Simultaneous X-ray/infrared (IR) observations in
August 1997 revealed such X-ray dips with corresponding large
amplitude IR flares, consistent with synchrotron emission from ejected
plasma bubbles (Eikenberry et al., 1998a - hereafter Paper 1).  Given
that GRS 1915+105 is known to eject synchrotron-emitting plasma at
relativistic speeds (Mirabel and Rodr\'{\i}guez, 1994), this leads to
a picture where the inner portion of an accretion disk is being swept
up and ejected from the system in the form of relativistic jets (Paper
1).  X-ray/IR/radio observations of dips/flares in September 1997
appear to confirm this interpretation (Mirabel et al., 1998).  Thus,
it seems that the microquasar GRS 1915+105 is giving the first
observational clues into the time-dependent interaction responsible
for producing relativistic jets from a black hole accretion disk.

	Despite these recent successes, we still know very little
about the binary system in GRS 1915+105.  Infrared spectra of this
heavily obscured object ($A_V \ \sim 27$ mag; Chaty et al., 1996) have
provided more controversy than conclusion.  Castro-Tirado et
al. (1996) observed variable, broad He\,I and H\,I emission lines from
GRS 1915+105.  Based on the apparent signature of a weak He\,II
emission line and the IR colors of the system (assuming $A_V \sim
18-24$ mag), they argued that the lines arise in an accretion disk
around the compact object and that GRS 1915+105 most likely has a
low-mass companion star.  Mirabel et al. (1997), on the other hand,
found that the IR spectra and colors of GRS 1915+105 (assuming $A_V
\sim 27$ mag) closely resemble those of many Oe/Be X-ray binaries,
where the broad emission lines arise in the circumstellar disk of the
high-mass companion star.  They failed to detect the He\,II line,
whose presence would argue against an Oe/Be companion since such stars
do not emit significant quantities of the 54 eV photons required to
doubly ionize helium.

	In this paper, we present a spectroscopic study of GRS
1915+105 during the same nights as the photometry presented in Paper
1, when the object exhibited both jet-producing and non-jet-producing
behaviors.  The results provide constraints for the origin of the IR
emission lines, the nature of the mass-donor companion in the binary
system, and the origin of the infrared flares accompanying the X-ray
dips.

\section{Observations and Data Reduction}

	We obtained infrared spectra of GRS 1915+105 using the HNA
longslit grating spectrometer (Larkin et al., 1996) in conjunction
with a $256 \times 256$ HgCdTe array camera on the Hale 200-inch
Telescope.  Conditions were photometric on all three nights, and
seeing was $\sim 0.5-0.8$ arcsec.  We tilted the spectrometer grating
in order to include the wavelengths of the known emission lines of
He\,I, He\,II, and H\,I from GRS 1915+105.  The wavelength coverage
was $\sim 0.12 \mu$m, centered near $\sim 2.1 \mu$m, with a spectral
resolution of $R \sim 875$.  We first took images of the source
through an open ($10 \times 40$ arcsec) and closed ($0.7 \times 40$
arcsec) slit with a flat mirror in front of the grating.  We then took
several 5-minute spectral exposures, displacing the source by
20-arcsec along the slit between exposures.  The telescope was
accurately positioned during these displacements using an image-motion
compensatory autoguider.  Repointing of the telescope took typically
$\sim 1$ minute, giving a final time resolution of about 6 minutes for
the spectra.  We also took spectra of bright G-dwarf stars at the same
grating tilts and roughly the same airmasses as GRS 1915+105 in order
to calibrate atmospheric absorption features.  We also obtained
simultaneous, intermittent X-ray observations using the Rossi X-ray
Timing Explorer (RXTE); details of these are provided in Paper 1.

	For each spectral exposure, we subtracted the next exposure,
removing bias and sky offsets.  After removing cosmic rays, we
corrected the frame for spatial distortions using a previously
measured transformation, and corrected for spectral distortions using
a quadratic fit to the telluric OH lines in the raw frames.  We then
divided each frame by a similarly processed G-star spectrum
(interpolating over the G-star's weak ${\rm Br} \gamma$ absorption
feature) and multiplied the result by a blackbody spectral shape
corresponding to the G-star's effective temperature to obtain the
final spectrum.

	Due to the photometric variability of GRS 1915+105 on these
nights (Paper 1), absolute calibration of the flux levels in the
spectra is difficult.  For Aug 13 UTC, imaging photometry showed that
the source had a fairly steady flux density level of $\sim 5$ mJy, and
we calibrated the Aug 13 spectra by defining this to be the continuum
flux density at $2.2 \mu$m.  Similarly, the minimum flux seen on Aug
14 and Aug 15 was $\sim 5$ mJy (Paper 1), and we used this to define
the {\it minimum} observed continuum flux density at $2.2 \mu$m on
these two nights.  Comparison of measured flux levels revealed a
repeated throughput variation in both the spectra and the slit images,
with the flux levels when the source is on one side of the slit
greater than those on the other side.  This was due to a misalignment
between the spectrometer slit and the telescope motions used to
position the source along the slit.  After correcting for this effect,
we estimate a $\pm 15$\% uncertainty in the flux levels due to the
systematic uncertainty in the placement of the source on the slit.

\section{Results}

	We present the dereddened, flux-calibrated, summed spectra for
each spectrograph grating tilt on each night in Figure 1.  All of the
spectra exhibit resolved, broad He\,I ($2.058 \mu$m) and Br$\gamma$
($2.1655 \mu$m) emission lines.  The longer wavelength spectrum taken
on Aug 14 shows a He\,II ($2.189 \mu$m) emission feature.  While the
signal-to-noise ratio for this feature is low ($\sim 4 \sigma$), its
properties (Table 1) closely match those seen by Castro-Tirado et
al. (1996).

	In addition to examining the summed spectra, we also searched
the individual 5-minute exposures for evidence of variability in both
the continuum flux densities and line fluxes.  We found significant
variability in the Aug 14 spectra.  Plotting a time series of the IR
continuum flux density level along with the X-ray count rate from the
RXTE PCA instrument at the same time (Figure 2), we surmise that we
observed portions of three infrared flares similar to those observed
photometrically on the same night (Paper 1).  We present individual
5-minute spectra taken during the first IR flare in Figure 3.  The
most striking feature in this figure is that the emission lines of
He\,I and H\,I are stronger in the high-flux spectra and weaker in the
low-flux spectra.  If this correlation was due to systematic
throughput variations (such as those we corrected for above), the line
equivalent width would remain constant.  However, when we examine the
individual equivalent widths, we see that they are {\it not} constant
(at the $99.2 \%$ confidence level).  Together with the similarities
in timescale, peak flux, and repetition period to the flares reported
in Paper 1, this confirms that the variations are intrinsic to GRS
1915+105.  When we plot the integrated Br$\gamma$ line flux versus the
continuum level (Figure 4), we see that they have an approximately
linear correlation, with a scatter corresponding to the variations in
the equivalent width.

\section{Discussion}

\subsection{The origin of the emission lines}

	As noted above, IR spectra of GRS 1915+105 are key to
understanding the nature of the binary system -- is it a high-mass or
low-mass system?  In the former case, the lines originate in the
circumstellar disk of an Oe/Be-type companion star, while in the
latter case, the lines originate in an accretion disk around the
compact object.  Since O and B stars do not produce the 54 eV photons
required to doubly ionize helium, the appearance of a weak He\,II line
in Figure 2 confirms the accretion disk scenario of Castro-Tirado et
al. (1996).  While we detect this line with a low signal-to-noise, the
fact that Castro-Tirado et al. reported a line with identical
properties indicates that this feature is most likely real.  The fact
that the other emission lines are manifestly time variable (Figures
3,4) may explain why the He\,II feature was not seen by Mirabel et
al. (1997) nor at other epochs by Castro-Tirado et al.  We also note
the similarity in the line ratios for He\,I, Br$\gamma$, and He\,II in
our spectra of GRS 1915+105 and the spectrum of Sco X-1 (Bandyopadhyay
et al., 1997), which is known to be a low-mass X-ray binary system
where the IR emission lines arise in an accretion disk surrounding the
compact object.  This similarity confirms that such accretion disks
are capable of producing the line characteristics we observe here.

	The nature of the emission line variability also seems to
support the accretion disk interpretation over the Oe/Be-star
interpretation.  The approximately linear correlation between the
emission line flux and continuum flux density in Figure 4 is a
signature of radiative pumping of the He\,I and H\,I emission lines by
the flares.  While it might be possible for a flare originating near
the compact object to pump the line-emitting region of a companion
star's circumstellar disk, roughly half of such a disk would be
shielded from the flare by the companion star itself.  Since the line
profile presumably results from the rotation of the disk, the
asymmetric illumination of the disk by the line-pumping flare should
dramatically change the line profile -- the increased flux should
originate from only one portion of the disk, causing the line FWHM to
decrease by a factor of $\sim 2$ and the line centroid to shift to the
blue or red as the line flux increases.  We can see from Figure 3 and
the measured line parameters (Table 1) that no such changes in the
line profile are evident.  In the accretion disk scenario, on the
other hand, the flare could uniformly illuminate the line-emitting
region of the disk, and no line profile changes would be evident,
which is consistent with the observations.

\subsection{The nature of the IR flares}

	In Paper 1, we demonstrated that the IR continuum emission
during the flares is probably dominated by synchrotron emission from
particles ejected from the system in the form of relativistic jets.
Meanwhile, the broadened profiles of the emission lines suggest that
they arise in a rotating disk.  However, the fact that the emission
line flux is proportional to the flare continuum flux density may
provide significant insights into the flares themselves.  The
proportionality suggests some sort of radiative pumping of the lines
by the flares, which in turn requires ultraviolet (UV) radiation
during the flares in order to ionize the H and He atoms producing the
line emission.  The required luminosity of ionizing radiation is only
$\sim 10^{25}$ W, but if the UV radiation originates in a distinct
region from the IR emission lines, this is only the fraction
intercepted by the line-emitting region, and thus represents a lower
limit to the total UV radiation during the flare.

	Another important insight into the IR flares comes from noting
the behavior of GRS 1915+105 on 13 Aug, when no IR flares were
observed.  On this night the Br$\gamma$ line flux was near the lowest
seen ($\sim 3 \times 10^{-17} {\rm W \ m^{-2}}$) while the X-rays were
in a high flux state ($L_x \ \sim 3 \times 10^{32}$ W).  Since these
X-rays arise as thermal emission in the inner accretion disk, they
will also be accompanied by thermal UV radiation.  Thus, the presence
of a large amount of ionizing radiation in the system is not
sufficient to produce a high emission-line flux.  This leads to two
key questions: (1) Why doesn't the high X-ray (and presumably UV) flux
on 13 Aug pump the IR emission lines?  (2) How does the presence of
flaring enable the pumping of the IR emission lines?

	Regarding the first question, as noted above both the X-ray
and IR line-emitting regions lie in the accretion disk plane.  For a
thin disk, essentially no ionizing radiation will reach the IR
emission region {\it through} the disk.  Therefore, unless the disk is
strongly curved, the IR line region will have a small solid angle as
seen from the inner disk, and will intercept little ionizing
radiation.  Thus, for a ``standard'' thin accretion disk, we may
expect that the inner disk radiation will not significantly pump IR
emission lines, which is consistent with the observations on 13 Aug
1997.  This leads to a possible answer to the second question: the
flares can enable the pumping of the IR emission lines due to the
presence of ejecta.  In Paper 1, we showed that the IR flares are
consistent with synchrotron emission from particles ejected from the
system.  If so, then as the particles move out of the disk plane, the
IR line-emitting region subtends a rapidly increasing solid angle as
seen from the ejecta.  Therefore, we suggest that IR emission line
variability may be due to radiative pumping by ionizing radiation from
the ejecta.  The ionizing radiation could be a UV tail to the
synchrotron emission from the ejected plasma which generates the IR
flare.

\section{Conclusions}

	We have presented near-IR medium-resolution ($R \sim 875$)
spectra of the Galactic microquasar GRS 1915+105 taken using the HNA
spectrometer on the Hale 200-inch telescope on 1997 Aug 13-15 UTC.
The spectra reveal broad emission lines of He\,I, H\,I ($\rm{Br}
\gamma$), and He\,II.  The presence of the He\,II line and the
relative constancy of the $\rm{Br} \gamma$ line profile as its flux
changes by factors of $\sim 5$ imply that the lines originate in an
accretion disk around the compact object, as suggested by
Castro-Tirado et al. (1996), rather than in the circumstellar disk of
an Oe/Be companion star, as suggested by Mirabel et al. (1997).  The
emission line flux is approximately linearly proportional to the
continuum flux during several infrared flares, implying that the
flares are associated with radiative pumping of the line emission.
This, together with the low line flux seen during a high X-ray flux
state on 13 Aug 1997, implies that the IR flares signify the ejection
of material out of the accretion disk plane, in agreement with the
results in Paper 1.

\acknowledgements The authors wish to thank G. Neugebauer for useful
discussions of these observations, and T. Prince for access to the
computers used to reduce this data.  SE acknowledges the support of a
Sherman Fairchild Postdoctoral Fellowship in Physics.  Infrared
astronomy at Caltech is supported by grants from NASA and the National
Science Foundation.

\begin{deluxetable} {lcccc}
\tablecolumns{5} 
\tablewidth{0pc} 
\tablecaption{Properties of emission lines observed from GRS 1915+105.
  In determining the line
FWHM, we have subtracted the spectrograph resolution ($\sim 340$ km/s)
in quadrature from the actual observed line width.  The centroid
uncertainty is for the last digit.  The first 4 rows
are for Br$\gamma$ lines in the summed spectra, the fifth row is for
the HeII line in the summed spectrum of 14 Aug, and the remaining rows
are for the Br$\gamma$ lines in the individual 300-second exposures on 14 Aug.  We estimate the uncertainties in the line properties by
first estimating the $\pm 1 \sigma$ noise in each pixel as the rms
residual for a low-order polynomial fit to a section of the continuum
spectrum.  Next, we add to each pixel a random number drawn from a
normal distribution with that $1 \sigma$ width, and re-measure the
line properties for this simulated spectrum.  Finally, we repeat this
process 1000 times, and take the standard deviation of the resulting
line properties as the $\pm 1 \sigma$ uncertainty,}
\tablehead{
\colhead{Date} & \colhead{UTC Time} &
\colhead{Flux} &
\colhead{Centroid} & \colhead{FWHM} \\
\colhead{} & \colhead{(hours)} &
\colhead{($10^{-18} \ {\rm W} \  {\rm m}^{-2}$)} &
\colhead{($\mu$m)} & \colhead{(${\rm km/s}$)}} 
\startdata 
13 Aug & 9.0-9.6 &  28 $\pm$6 & 2.1644 $\pm$8 & 360 $\pm$190 \nl
14 Aug & 8.4-8.9 &  66 $\pm$12 & 2.1651 $\pm$4 & 230 $\pm$60 \nl
14 Aug & 8.9-9.6 &  65 $\pm$11 & 2.1650 $\pm$4 & 330 $\pm$80 \nl
15 Aug & 8.7-9.2 &  28 $\pm$7 & 2.1640 $\pm$20 & 230 $\pm$300 \nl
\nl
14 Aug & 8.9-9.6 & 15 $\pm$5 & 2.189 $\pm$12 & 440 $\pm$210 \nl
\nl
14 Aug (A)& 8.4-8.5 &  30 $\pm$11 & 2.1663 $\pm$7 & 640 $\pm$220 \nl
14 Aug (B)& 8.5-8.6 &  115 $\pm$23 & 2.1649 $\pm$7 & 230 $\pm$110 \nl
14 Aug (C)& 8.6-8.7 &  64 $\pm$15 & 2.1650 $\pm$6 & 200 $\pm$200 \nl
14 Aug (D)& 8.7-8.8 &  97 $\pm$25 & 2.1653 $\pm$4 & 360 $\pm$220 \nl
14 Aug (E)& 8.8-8.9 &  25 $\pm$10 & 2.1654 $\pm$17 & 200 $\pm$200 \nl
14 Aug & 8.9-9.0 &  50 $\pm$12 & 2.1643 $\pm$9 & 330 $\pm$120 \nl
14 Aug & 9.0-9.1 &  134 $\pm$28 & 2.1655 $\pm$4 & 200 $\pm$200 \nl
14 Aug & 9.1-9.2 &  \nodata & 2.1646 $\pm$4 & 200 $\pm$200 \nl
14 Aug & 9.2-9.3 &  55 $\pm$12 & 2.1644 $\pm$6 & 200 $\pm$140 \nl
14 Aug & 9.3-9.4 &  60 $\pm$17 & 2.1641 $\pm$40 & 200 $\pm$250 \nl
14 Aug & 9.4-9.5 &  24 $\pm$11 & 2.1658 $\pm$6 & 200 $\pm$280 \nl
14 Aug & 9.5-9.6 &  70 $\pm$18 & 2.1652 $\pm$9 & 550 $\pm$200 \nl
\enddata

\end{deluxetable}

\begin{figure}
\vspace*{180mm}
\includegraphics{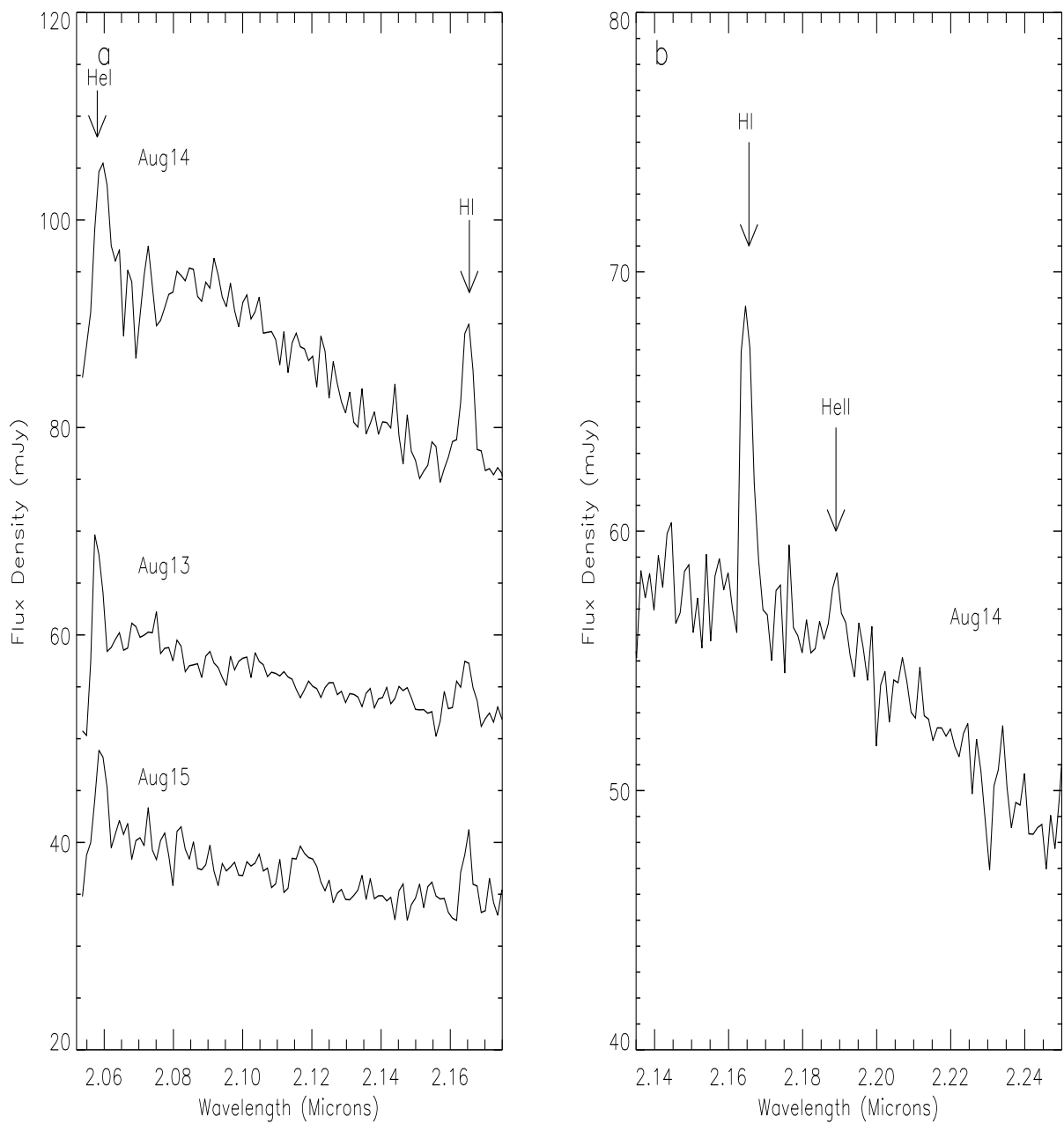}
\caption{Summed infrared spectra from GRS 1915+105 over two wavelength
intervals.  In (a), the Aug 13 spectrum has been offset by $+15 \ {\rm
mJy}$, for the sake of clarity.  Arrows indicate emission lines of
He\,I ($2.058 \mu$m), HI (Br$\gamma$ = $2.166 \mu$m), and He\,II
($2.189 \mu$m).  Unfortunately, the He\,I line, while obviously
present, is ``falling off'' the detector array, making quantitative
measurements of its properties difficult.  All spectra have been
dereddened assuming $A_V = 27$ mag ($A_K = 3.0$ mag)}
\end{figure}

\begin{figure}
\vspace*{180mm}
\includegraphics{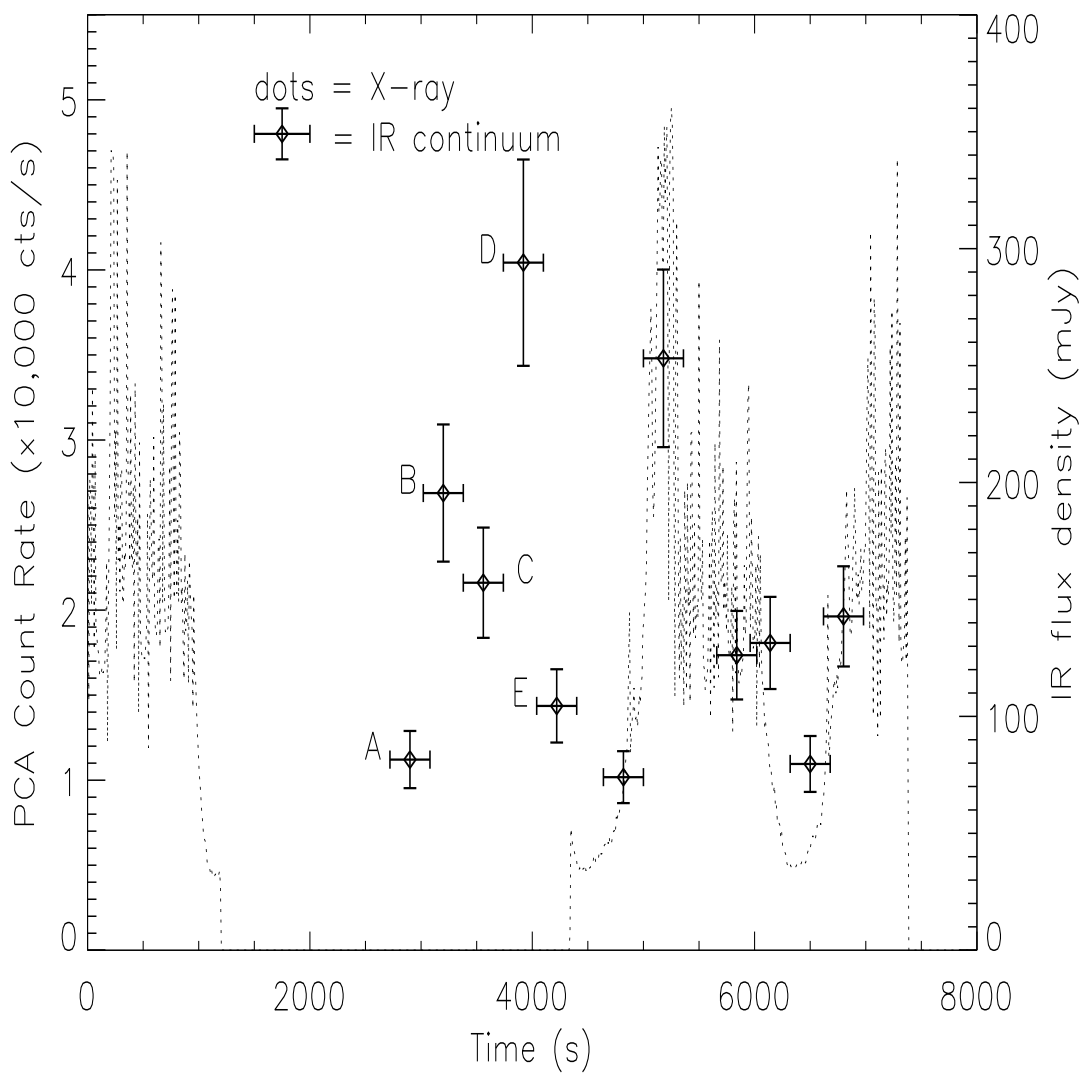}
\caption{IR continuum flux density at $2.16 \mu$m and RXTE PCA X-ray
count rate versus time on 1997 Aug 14 UTC.  The X-ray count rates have
been averaged over 8-second intervals, and the gaps between $\sim 1200
- 4300$ s and after $\sim 7300$s are due to Earth occultations.  The
IR flux densities have been dereddened assuming $A_V = 27$ mag ($A_K =
3.0$ mag).  Error bars indicate the $\pm 1 \sigma$ uncertainties.  The
origin of the time axis corresponds to 7:36 UT.}
\end{figure}

\begin{figure}
\vspace*{180mm}
\includegraphics{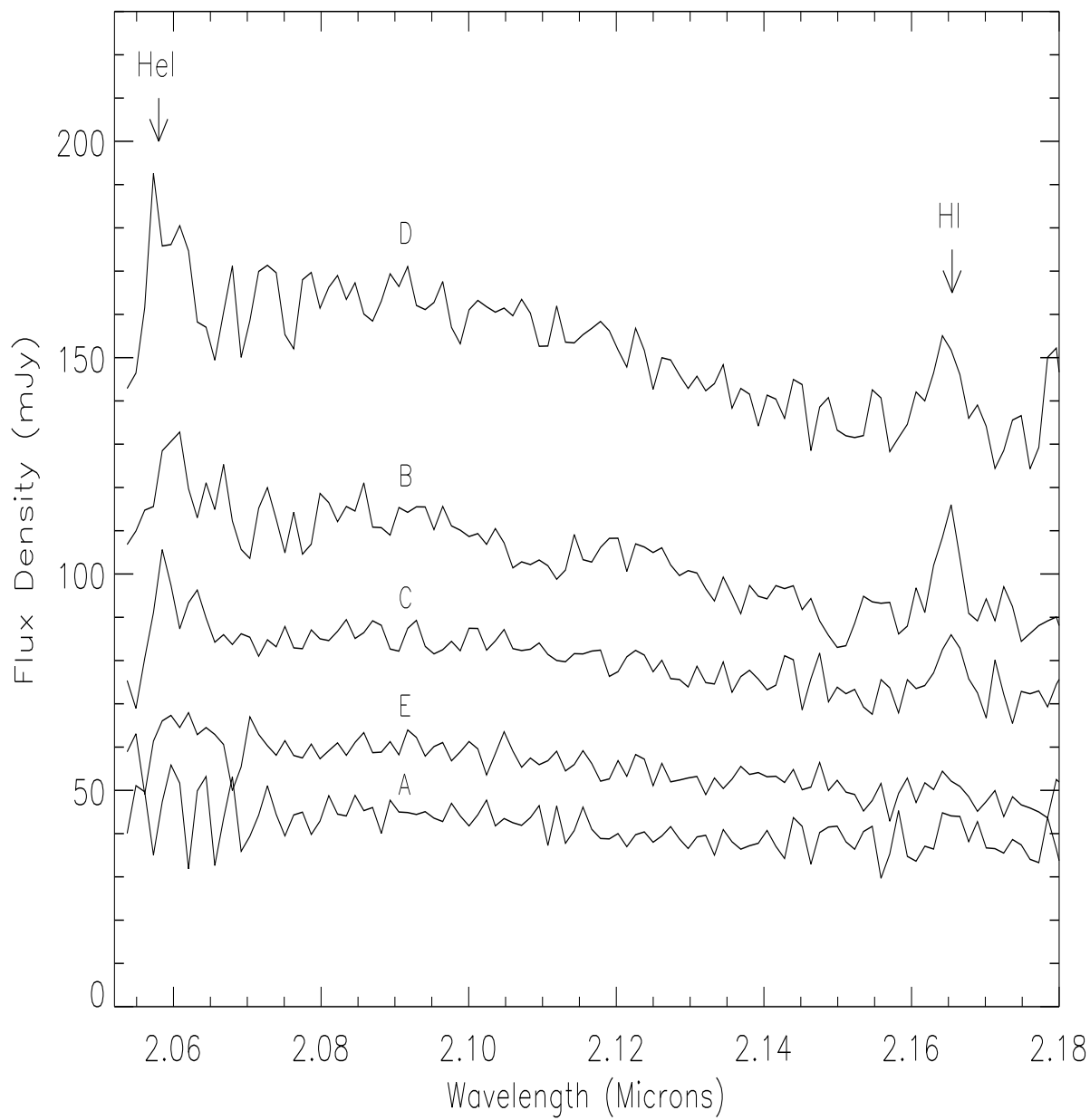}
\caption{Individual 5-minute spectral exposures during an IR flare on
1997 Aug 14 UTC.  The time order of the spectra is A,B,C,D,E (see
Figure 2).  Note that the overall offset of B above C is consistent
with 0 within the photometric uncertainties.  Arrows indicate emission
lines of He\,I ($2.058 \mu$m) and HI (Br$\gamma \ = \ 2.166 \mu$m).
Note that the emission line strength appears to be correlated with
continuum flux level.  All spectra have been dereddened assuming $A_V
= 27$ mag ($A_K = 3.0$ mag).}
\end{figure}


\begin{figure}
\vspace*{180mm}
\includegraphics{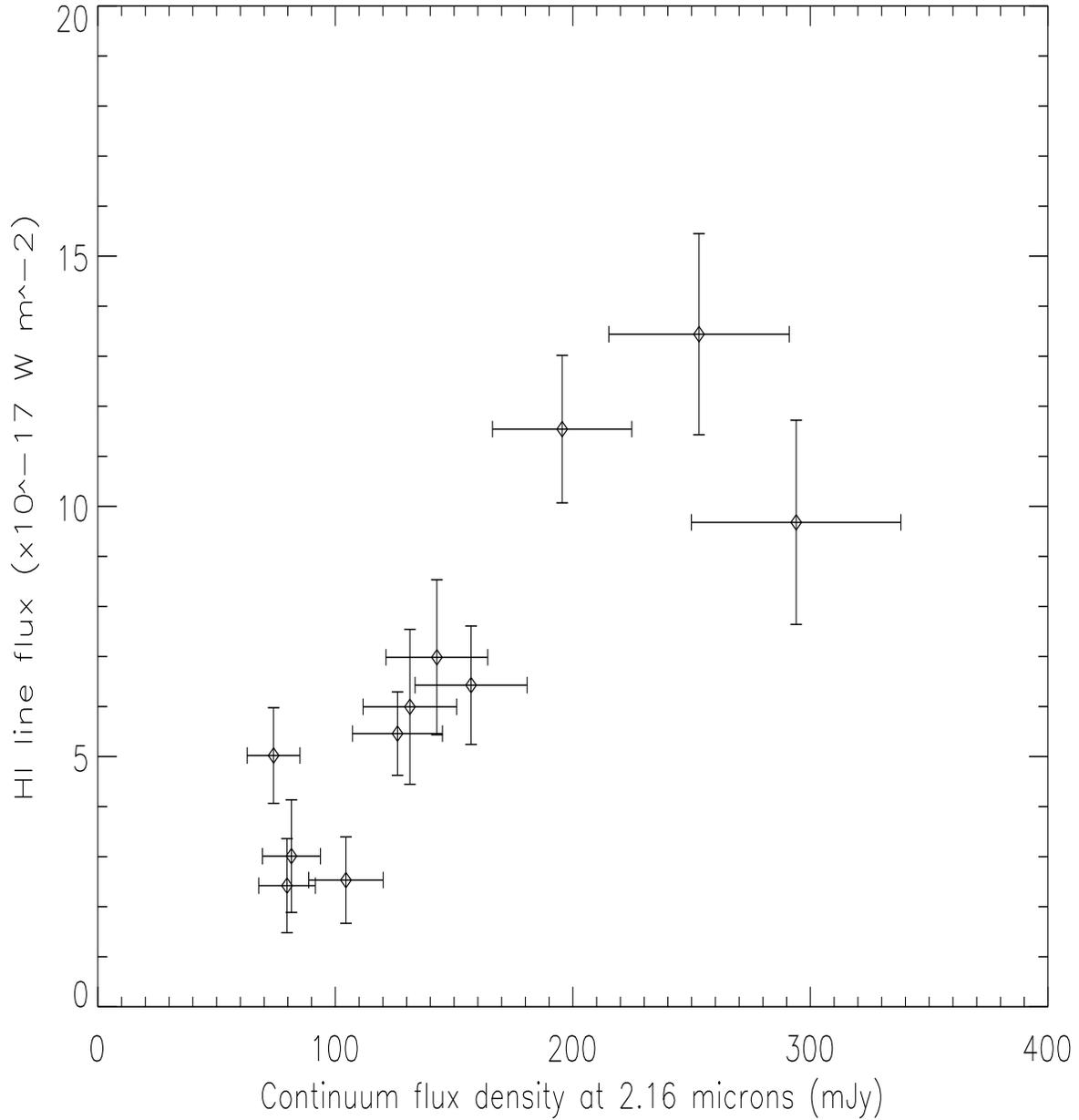}
\caption{HI (Br$\gamma \ = \ 2.166 \mu$m) integrated line flux versus
$2.16 \mu$m continuum flux density for spectra taken during IR flares
on 1997 Aug 14 UTC.  Error bars indicate the $\pm 1 \sigma$
uncertainties.  Note the approximately linear dependence of the line
flux on the continuum flux density.  All spectra have been dereddened
assuming $A_V = 27$ mag ($A_K = 3.0$ mag).}
\end{figure}

\end{document}